\documentstyle[aps,epsfig,pnas,fancyheadings]{revtex}

\newcommand {\be   } {\begin{equation}}
\newcommand {\ee   } {\end{equation}  }

\begin{document}

\title{How native state topology affects the folding of
Dihydrofolate Reductase and Interleukin-1$\beta$}

\author{Cecilia Clementi$^{1}$, Patricia A. Jennings$^{2}$,
Jos\'e N. Onuchic$^{1}$}

\address{Department of Physics$^{1}$ and        
Department of Chemistry and Biochemistry$^{2}$ }
\address{University of California at San Diego
La Jolla, California 92093, USA}

\maketitle





\vspace{1cm}


\begin{abstract}
The overall structure of the transition state and intermediate
ensembles experimentally observed for {\it Dihydrofolate Reductase}
and {\it Interleukin}-1$\beta$ can be obtained utilizing simplified
models which have almost no energetic frustration. The predictive
power of these models suggest that, even for these very large proteins
with completely different folding mechanisms and functions, real
protein sequences are sufficiently well designed and much of the
structural heterogeneity observed in the intermediates and the 
transition state ensembles is determined by topological effects.
\end{abstract}

\newpage
\twocolumn

\section{Introduction}

Explaining how proteins self-assemble into well defined
structures is a longstanding challenge. Energy landscape theory and
the funnel concept
\cite{Leopold92,Onuchic97,Dill97,Nymeyer98,Klimov96,Shakhnovich96:FD,Shea98}
have provided the theoretical framework necessary for improving our
understanding of this problem --- efficient folding sequences
minimize frustration. Frustration may arise from the inability to
satisfy all native interactions and from strong
non-native contacts which can create conformational traps. The difficulty
of minimizing energetic frustration by sequence design, however, is
also dependent on the choice of folding motif.  Some folding motifs
are easier to design than others \cite{LiH96,Nelson98}, suggesting the
possibility that evolution not only selected sequences with
sufficiently small energetic frustration but also selected more easily
designable native structures. To address this difference in
foldability, we have introduced the concept of ``topological
frustration'' \cite{Nymeyer99:PNAS,Onuchic99,Shea99,Clementi99} --- even when
sequences have been designed with minimal energetic frustration,
variations in the degree of nativeness of contacts in the transition
state ensemble (TSE) are observed because of asymmetries imposed by
the chosen final structure. 


Recent theoretical and experimental evidences suggest that proteins,
especially small fast folding (sub-millisecond) proteins, have
sequences with a sufficiently reduced level of energetic frustration
that the global characteristics of the observed heterogeneity observed
in the TSE are strongly influenced by the native state topology. We
have shown \cite{Clementi99} that the overall structure of the TSE for
{\it Chymotrypsin Inhibitor 2} (CI2) and for the SH3 domain of the
{\it src tyrosine-protein kinease} can be obtained by using simplified
models constructed by using sequences that have almost no energetic
frustration (G\=o--like potentials). These models drastically reduce
the energetic frustration and energetic heterogeneity for native
contacts, leaving the topology as the primary source of the residual
frustration. Topological effects, however, go beyond affecting the
structure of the TSE.  The overall structure of the populated
intermediate state ensembles during the folding of proteins
such as {\it Barnase}, {\it Ribonuclease H} and {\it CheY} have also 
been successfully determined using a similar model \cite{Clementi99}. 
It is interesting to notice that although these model, since they consider
totally unfrustrated sequences, may not reproduce the precise energetics 
of the real proteins, such as the value of the barrier heights and the 
stability of the intermediates, they are able to determine the general 
structure of these
ensembles. Therefore, the fact that these almost energetically
unfrustrated models reproduce most of the major features of the TSE of
these proteins indicate that real protein sequences are sufficiently
well designed (i.e. with reduced energetic frustration) that much of
the heterogeneity observed in the TSE's and intermediates have a
strong topological dependence.

Do these conclusions hold to larger and slower folding proteins with a
more complex folding kinetics than two--state folders as CI2 and SH3?
The success obtained with {\it Barnase}, {\it Ribonuclease H} and {\it CheY}
intermediates already provides
some encouragement --- topology appears to be important in determining
on-pathway folding intermediates. In this paper this approach is
extended to a pair of larger proteins: {\it Dihydrofolate Reductase}
(DHFR) and {\it Interleukin}-1$\beta$ (IL-1$\beta$). The synoptic
analysis of these two proteins is particularly interesting because
they have a comparable size (slightly over 150 amino--acids), but
different native structures, folding mechanisms and functions: DHFR is
a two--domain $\alpha$/$\beta$ enzyme that maintains pools of
tetrahydrofolate used in nucleotide metabolism while IL-1$\beta$ is a
single domain all $\beta$ cytokine with no catalytic activity on its
own but elicits a biological response by binding to its receptor.

\section{Numerical procedures}
\label{num.proc:sec}
  
The energetically unfrustrated model of DHFR and IL-1$\beta$ are
constructed by using a G\=o--like Hamiltonian \cite{Ueda75,Ueda78}.
A G\=o--like potential takes into account only native interactions, 
and each of these interactions enters in the energy balance
with the same weight.  Residues in the proteins are represented as 
single beads centered in their C--$\alpha$ positions. 
Adjacent beads are strung together into a polymer chain by means of 
bond and angle interactions, while the geometry of the native state 
is encoded in the dihedral angle potential and a non-local bead-bead 
potential.

A detailed description of this energy function can be found elsewhere
\cite{Clementi99}.  The local (torsion) and non-local terms have been
adjust so that the stabilization energy residing in the tertiary
contacts is approximately twice as large as the torsional
contribution.
This balance among the energy terms is optimal for the folding of our
G\=o--like protein models \cite{Nymeyer98}.  
Solvent mediation and side chain effects are
already included in these effective energy functions. Therefore, entropy changes
are associated to the configurational entropy of the chain. 
The native contact map of
a protein is derived with the CSU software, based upon the approach
developed in ref. \cite{Sobolev96}. Native contacts between pairs of
residues $(i,j)$ with $j \leq i+4$ are discarded from the native map
as any three and four subsequent residues are already interacting in
the angle and dihedral terms. A contact between two residues $(i,j)$
is considered formed if the distance between the $C_{\alpha}$s is
shorter than $\gamma$ times their native distance $\sigma_{ij}$. It
has been shown \cite{Onuchic99} that the results are not strongly
dependent on the choice made for the cut--off distance $\gamma$. In
this work we used $\gamma = 1.2$.

For both (DHFR and IL-1$\beta$) protein models, folding and unfolding
simulations have been performed at several temperatures around the
folding temperature.  The results from the different simulations have
been combined using the WHAM algorithm \cite{Swendsen93}.  Several
very different initial unfolded structures for the folding simulations
have been selected and they have been obtained from high temperature
unfolding simulations. 
In order to have appropriate
statistics, we made sure that for every transition state ensemble or
intermediate, we have sampled about 500 uncorrelated conformations
(thermally weighted). For smaller proteins such as SH3 and CI2 (that
have about 1/3 of the tertiary contacts of DHFR and 
IL-1$\beta$) we have determined that about 200 uncorrelated
conformations in the transition state ensemble are necessary to have
an error on the estimates of contact probabilities (or $\Phi$ values)
of $\pm \ 0.05$ \cite{Clementi99}.

\section{Comparing simulations and experiments for Dihydrofolate Reductase
and Interleukin--1$\beta$}

{\it Dihydrofolate Reductase} and {\it Interleukin}-1$\beta$ not only
have dissimilar native folds\footnote{The 162 residues of DHFR
arrange themselves in 8 $\beta$-strands and 4 $\alpha$-helices,
grouped together in the folded state in as detailed in
Fig.\ref{fig3:fig}~(d), while IL-1$\beta$ is a 153 residues,
all-$\beta$ protein, composed by 12 $\beta$ strands packed together as
shown in Fig.\ref{fig4:fig}~(c)-(d).} but also the nature of the
intermediate states populated during the folding event is remarkably
different. To explore the connection between the protein topology and
the nature of the intermediates, we used an energetically minimally
frustrated $C_{\alpha}$ model for these two proteins, with a potential
energy function defined by considering only the native local and
non-local interactions as being attractive (see Numerical Procedures,
for details). This is a very simplified potential that retains only
information about the native fold --- energetic frustration
is almost fully removed.
Notice that although the real amino--acid sequence
is not included in this model, the chosen potential is like a
``perfect'' sequence for the target structure, without the energetic
frustration of real sequences (since this potential includes
attractive native tertiary contacts, it implicitly incorporates
hydrophobic interactions). Therefore,  this model provide us with the
perfect computational tool to investigate how much of the structural
heterogeneity observed during folding mechanism could be inferred from
the knowledge of the native structure alone without contributions from
energetic frustration. 

Since early work suggests that proteins (at least small fast folding
proteins) have sufficiently reduced energetic frustration, they have
a funnel-like energy landscape with a solvent--averaged potential
strongly correlated with the degree of nativeness (but with some
roughness due to the residual frustration). In this situation, the
folding dynamics can be described as the diffusion of an ensemble of
protein configurations over a low dimensional free energy surface ---
defined in terms of the reaction coordinate $Q$, where $Q$ represents
the fraction of the native contact formed in a conformation ($Q=0$ at
the fully unfolded state and $Q=1$ at the folded state)
\cite{Nymeyer99:PNAS,Onuchic99,Shea99,Clementi99,Onuchic96}. The ensemble of
intermediates observed in this free energy profile are expected to mimic
the real kinetic intermediates.

Fig. \ref{fig1:fig} shows a comparison between the folding mechanism
obtained from our simulations for the minimally-frustrated analogue of
DHFR (panels (a) and (c)) and IL-1$\beta$ (panels (b) and (d)).  The
different nature of the folding intermediates of the two proteins and
their native ensembles emerging from these data is in substantial
agreement with the experimental observations, with the adenine binding
domain of DHFR being folded in the main intermediate in the simulation
and the central $\beta$ strands of IL-1$\beta$ being formed early in
this single domain protein.  The absolute values of the free energy barriers
resulting from simulations may not necessarily agree with the
experimental ones because we are dealing with unfrustrated designed
sequences.  
Thus, quantitative predictions that
depend on barrier heights and stability of the intermediate ensembles
(e.g.  folding time, rate determining barriers and lifetime of
intermediates) are not possible for this kind of models. 
However we show that topology is sufficient to correctly detect the
positions of the transition state and intermediate states.  A more
detailed description follows.

\subsection{Dihydrofolate Reductase}
 
The folding process emerging from the dynamics of the G\=o--like
analogue of DHFR (as summarized in Fig. \ref{fig1:fig}~(a) and (c))
is interestingly peculiar and consistent with the experimentally
proposed folding mechanism \cite{Jennings93a} (see Fig.
\ref{fig2:fig}~(d)). Refolding initiates by a barrierless collapse to
a quasi--stable species (Q=0.2) which corresponds to the formation of
a burst--phase intermediate, $I_{BP}$, with little stability but some
protection from $H$-exchange across the central $\beta$ 
sheet~\cite{Jones95}. This
initial collapse is followed by production of the main intermediate
$I_{HF}$ {\it (Highly Fluorescent)}, which is described in the 
mechanism of Fig.  \ref{fig2:fig}~(d) as the collection of intermediates $I_1 - I_4$.
$I_1 - I_4$ are structurally similar to each other, but differentiated
experimentally by the rate at which they proceed towards the native protein.
Finally, after the overcoming of a second barrier, the protein visits
an ensemble of native structures with different energies. The
experimentally determined folding mechanism of DHFR shows
transient kinetic control in the formation of native conformers
($N_4$ dominant). This is later overridden by thermodynamic
considerations ($N_2$ dominant) at final equilibrium \cite{Jennings93a}.
This latter finding is consistent with the nature of the folding ensemble 
determined by the simulations. As shown in Fig. \ref{fig2:fig}~(b) 
a set of structures close to to the native state (Q around 0.7-0.8) is
transiently populated beside the fully folded state (Q = 1). 
Since the main intermediate $I_{HF}$ has been recently
characterized by experimental studies 
\cite{Jennings93a,Jones95,Jones94,Kuwajima91,Jones95b,Heidary99}, we take our
analysis a step farther by comparing the average structure of
the $I_{HF}$ ensemble from our simulations to the one experimentally
determined. For this purpose we compute the formation probability
$Q_{ij}(Q)$ for each native DHFR contact --involving residues
$(i,j)$-- at different stages of the folding process by averaging the
number of times the contact occurs over the set of structures
existent in a selected range of $Q$. As detailed in
Fig.~\ref{fig3:fig}, the central result from this analysis is that
the main intermediate $I_{HF}$ is characterized by a largely
different degree of formation in different parts of the protein: {\it
domain 1} (i.e. interactions among strands 2-5 and helices 2-3)
appears to be formed with probability greater than 0.7 while {\it
domain 2} (i.e. interactions among strands 6-8, helix 1 and helix 4)
is almost non existent.

The formation of {\it domain 1} and {\it domain 2} during the folding
event is more closely understood from Fig.~\ref{fig2:fig} 
(panels (a) and (c)), where the
RMS distance of the parts of the protein constituting each {\it
domain} from the corresponding native structures is shown for a
typical folding simulation.  Indeed the two {\it domains} fold in a
noticeably different way: in the stable intermediate $I_{HF}$,
{\it domain 1} is closer than 5 \AA ~(RMS) to that found in the 
native structure
while {\it domain 2} is highly variable (RMS distance
greater than 15 \AA ~from its native structure).  
Still, in agreement with hydrogen exchange studies \cite{Jones95}, some
protection is expected across domains from our simulations and complete
protection from exchange is expected only after the formation of 
the fully folded protein.  A combination of
fluorescence, CD mutagenic and new drug binding studies on DHFR indeed
demonstrate that {\it domain 1} is largely folded with specific tertiary
contacts formed and that this collection of intermediates is
obligatory in the folding route \cite{Heidary99}.

\subsection{Interleukin--1$\beta$}

Supported by some recent experiments, Heidary et al \cite{Heidary97}
have proposed a kinetic mechanism for the folding of IL-1$\beta$ that
requires the presence of a well defined on-pathway intermediate
species. The structural details of these species were determined from
NMR and hydrogen exchange techniques \cite{Heidary97,VarleyP93}.
We have compared these experimental data with our simulations 
for the IL-1$\beta$ G\=o--like analogue ( Fig. \ref{fig1:fig}~(b) and (d)).
The folding picture emerging from these numerical studies differs
substantially from that observed for  DHFR (see panels (a) and (c) of
Fig. \ref{fig1:fig}).  An intermediate state is populated for $Q$
around 0.55, followed by a rate limiting barrier (around $Q=0.7$)
after which the system proceeds to the well defined native state.

Is the theoretical intermediate similar to the one observed 
experimentally? Using the same procedure employed for the DHFR, a 
comparison between the average structure of the IL-1$\beta$  
intermediate ensemble and the one emerging from 
experimental studies is shown in Fig. \ref{fig4:fig}.
These results indicate that the
calculated intermediate has residues 40-105 (strands 4-8) folded into
a native-like topology but with interactions between strands 5 and 8 
not fully completed. 
Experimental results confirm that strands 6-8 are well folded in the
intermediate state and that strands 4-5 are partially formed. 
However results of experiments and theory differ in
the region between residues 110-125 where hydrogen exchange shows 
early protection and theory predicts late contact formation.
This region contains 4 aromatic groups PHE 112, PHE 117, TYR 120, TRP 121
which may be sequestered from solvent due to clustering of these
residues and from removal from unfavorable solvent interactions. 
This effect would not be fully accounted for our model, where all
native interactions are considered as energetically equivalent and
large stabilizing interactions are not differentiated.
Thus, energetics may favor early formation of the structure corresponding
to residues 105-125 while topology considerations favor the formation
of strands 4-8.
  
\section{Conclusions}

Theoretical and experimental studies of protein folding at times
appear to be at odds. Theoretical analysis of simple model systems
oftentimes predict a large number of routes to the native protein
whereas experimental work on larger systems indicates that folding
proceeds through a limited number of intermediate species. Although 
in the eyes of some people these
two descriptions are inconsistent with each other, this is
clearly not true. The large number of routes may or may not lead to
the production of on-route kinetic intermediate ensembles depending on the
result of the competition between configurational entropy and the
effective folding energy. In this study, we show that productive
intermediate species are produced by using simplified protein models,
with funnel-like landscapes, based on purely topological
considerations and the results are in good agreement with the
available experimental data.  The fact that these simplified minimally
frustrated models for DHFR and IL-1$\beta$ can predict the overall
features of the folding intermediates and transition states
experimentally measured for these two proteins, with completely
different folding mechanisms and functions, support our general picture
that real proteins have a substantially reduced level of energetic
frustration and a large component of the observed heterogeneity during
the folding event is topologically determined. Such observations lead
us to propose that the success in designing sequences that fold to a
particular shape is constrained by topological effects.  What is more
challenging are the consequences of this conclusion --- are these
topological constraints something that only have to be tolerated
during the folding event or are they actually used by biology towards
helping function?  Here we speculate only in the context of these two
examples, but this question really should be addressed more generally
in the future.

\section{acknowledgments}

This work has been supported by the NSF (Grant \# 96-03839), the 
La Jolla Interfaces in Science program (sponsored by the Burroughs 
Wellcome Fund), and the NIH (grant \# 6M54038).
We warmly thank Angel Garc\'{\i}a for many fruitful discussions.
One of us (C.C.) expresses her gratitude to Giovanni Fossati for his 
suggestions and helpful discussions.

\newpage


\newpage

\centerline{FIGURE CAPTIONS}

{\bf Fig. 1}
{\bf (a)} RMS distances between the DHFR native structure and several
computationally determined structures at different values of the
reaction coordinate $Q$ for an unfolding simulation at a temperature
slightly above the folding temperature ($T = 1.01 T_f$) and {\bf (c)}
free energy $F(Q)$ of the DHFR G\=o--like model as a function of $Q$
around the folding temperature.  The folding temperature $T_f$ is
estimated as the temperature where a sharp peak appears in the
specific heat plotted as a function of the temperature (data not
shown). Both temperatures and free energies are presented in units of
$T_f$.  
Notice that the thermal fluctuations around the
lowest energy state (i.e. $Q =1$, by construction of the model)
account for motions around the free energy minimum. Therefore, the
folded state ensemble has a minimum close to $Q =1$,at $Q \sim 0.9$,
but not exactly at $Q=1$. Indeed at $Q=1$ the structure would be
frozen in the native configuration. A similar remark applies for the
IL-1$\beta$ free energy profile shown in panel (c). 
The energy of a configuration, as quantified by the
color scale on the top of the figure, is here defined as the bare
value of the effective potential function in that configuration
(i.e. no configurational entropy is accounted in the
energy). Differences between energy and free energy (at finite
temperature) are due to the configurational entropy contribution to
the free energy. 
In panel (c) a main intermediate ensemble $I_{HF}$ emerges in the
folding process as a local minimum at $Q$ around 0.4 after the
overcoming of the first barrier.  Indeed this local minimum
corresponds to a populated region in panel (a) (after the scarcely
populated barrier around $Q$ = 0.3) with energy significantly lower
than in the unfolded state.  This main intermediate then evolves
toward a set of structures close to the native states (located between
$Q=0.7$ and $Q=0.8$) that eventually interconvert into the fully
folded state. A transient set of structures, close to the native
state, is also apparent from Fig.~\protect\ref{fig2:fig}~(b).  The
folding scheme resulting from these simulations is consistent with the
sketch of Fig.~\protect\ref{fig2:fig}~(d), proposed from the
experimental data \protect\cite{Jennings93a,Heidary99}.

{\bf (b)} RMS distances between the native structure and several
computationally determined structures at different values of the
reaction coordinate $Q$ for a folding simulation of the G\=o--like
model of IL-1$\beta$. The simulation is performed at a temperature
near to the folding temperature ($T = 0.99 T_f$).  {\bf (d)} Free
energy $F(Q)$ as a function of $Q$ around to the folding
temperature. The folding temperature is estimated from the sharp peak
in the specific heat curve as a function of the temperature (data not
shown). An intermediate ensemble is populated during the folding event
and it is identified by the broad local minimum in the free energy
profile (around $Q$ = 0.55), and the corresponding populated region in
panel (b) (with energy significantly lower than in the unfolded
state). These results are consistent with the kinetic mechanism for
the folding of IL-1$\beta$ proposed by Heidary et
al. \protect\cite{Heidary97}.  A set of structures close to the native
conformation is transiently populated for $Q$ between 0.75 and 0.8
(see panel (b) and the corresponding ``flat" region in the free energy
panel (d) ). This fact could be interpreted as the presence of an
additional intermediate state close to the native
state. Experimentally the possibility that another partially unfolded
form could be populated during the folding process is currently under
investigation.  Several constant temperature simulations 
(both folding and unfolding simulations) 
of the two protein models were made and combined to generate the free
energy plots.

\begin{figure}[h]
\vspace{2cm} 
\centerline{\psfig{figure=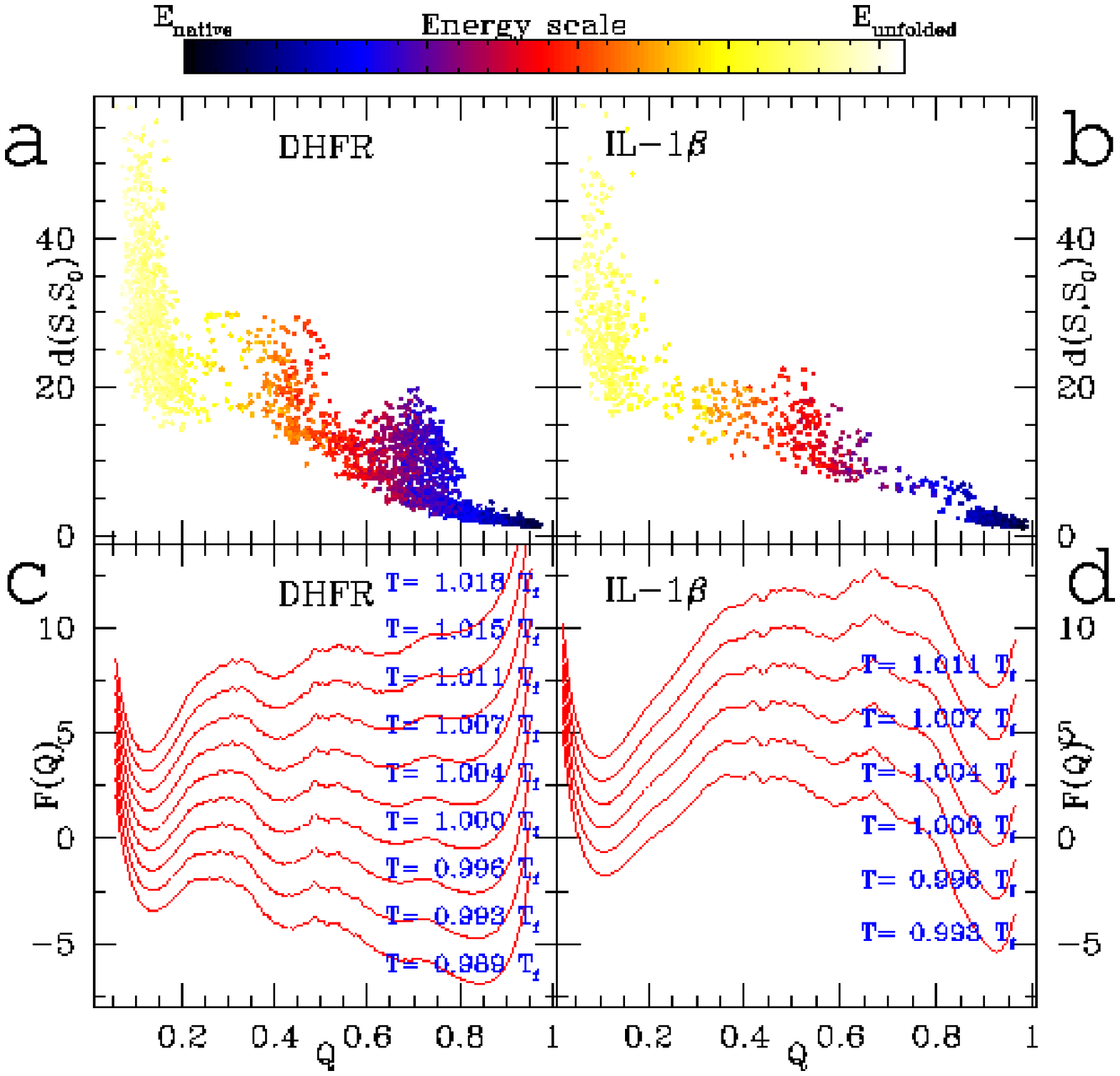,width=0.9\textwidth,clip=}}
\caption{
}
\label{fig1:fig}
\end{figure}

\begin{figure}[h]
\vspace{2cm} 
\centerline{\psfig{figure=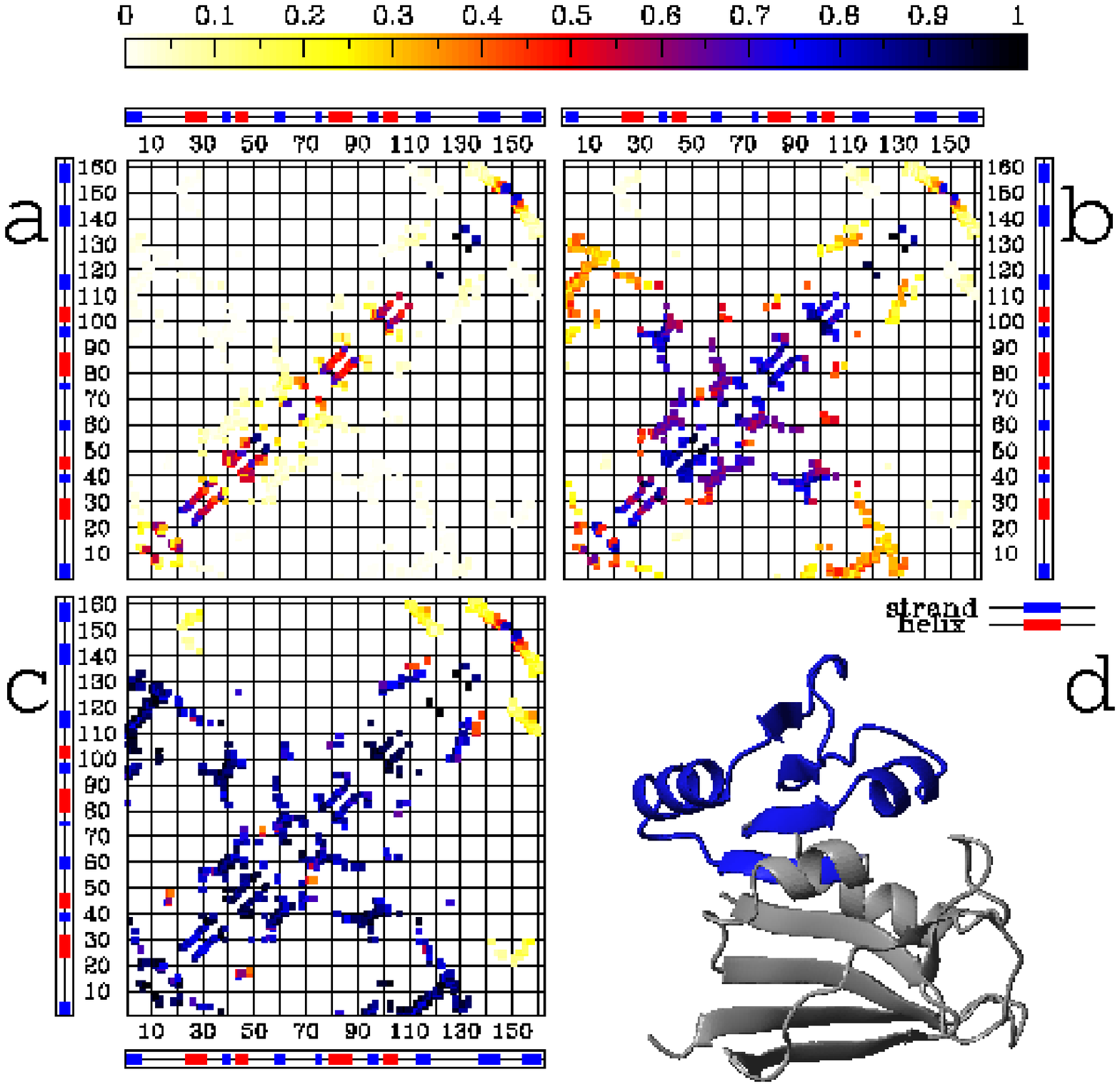,width=0.9\textwidth,clip=}}
\caption{
}
\label{fig3:fig}
\end{figure}

\begin{figure}[h]
\vspace{2cm} 
\centerline{\psfig{figure=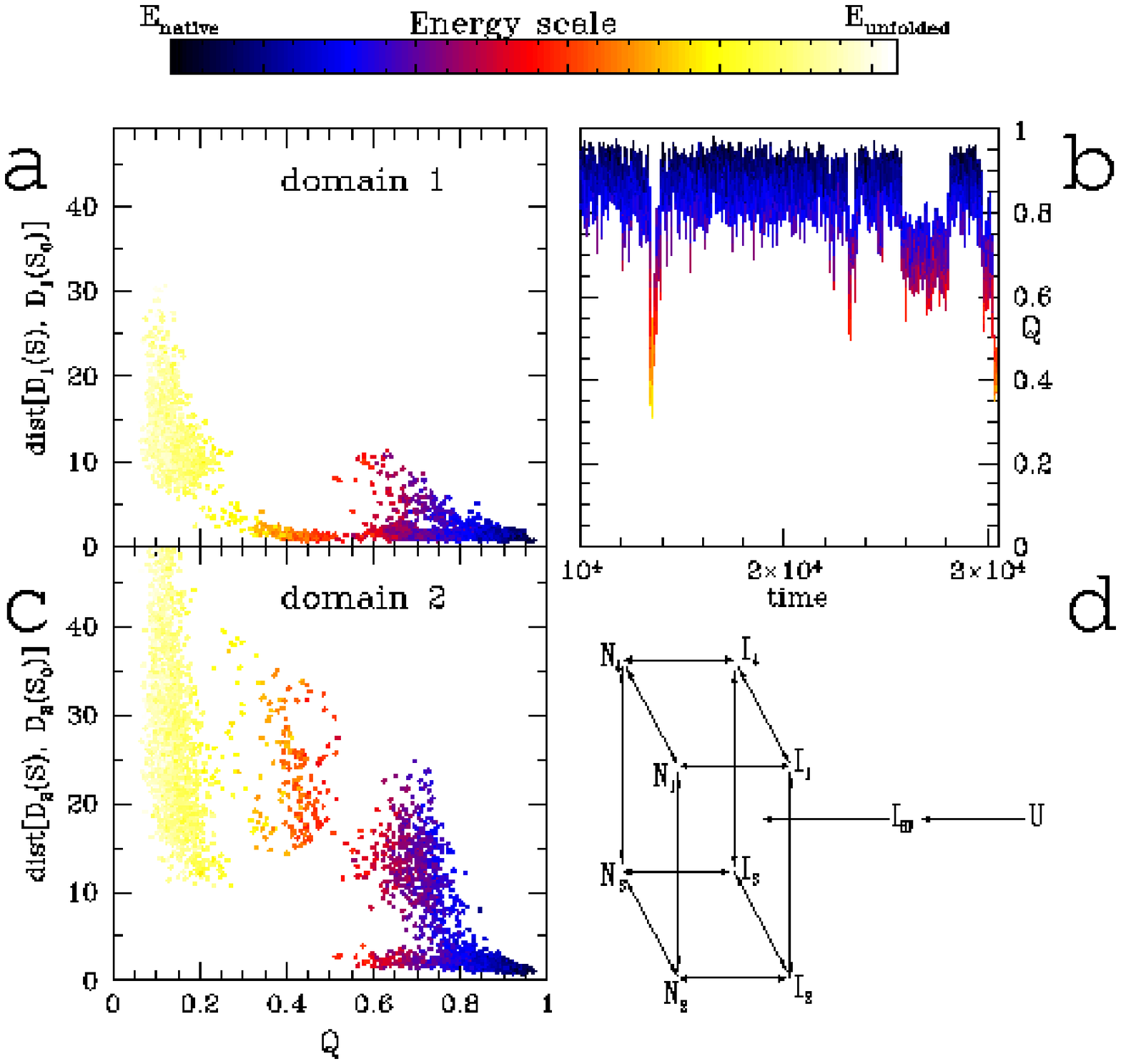,width=0.9\textwidth,clip=}}
\caption{
}
\label{fig2:fig}
\end{figure}

\begin{figure}[h]
\vspace{2cm}
\centerline{\psfig{figure=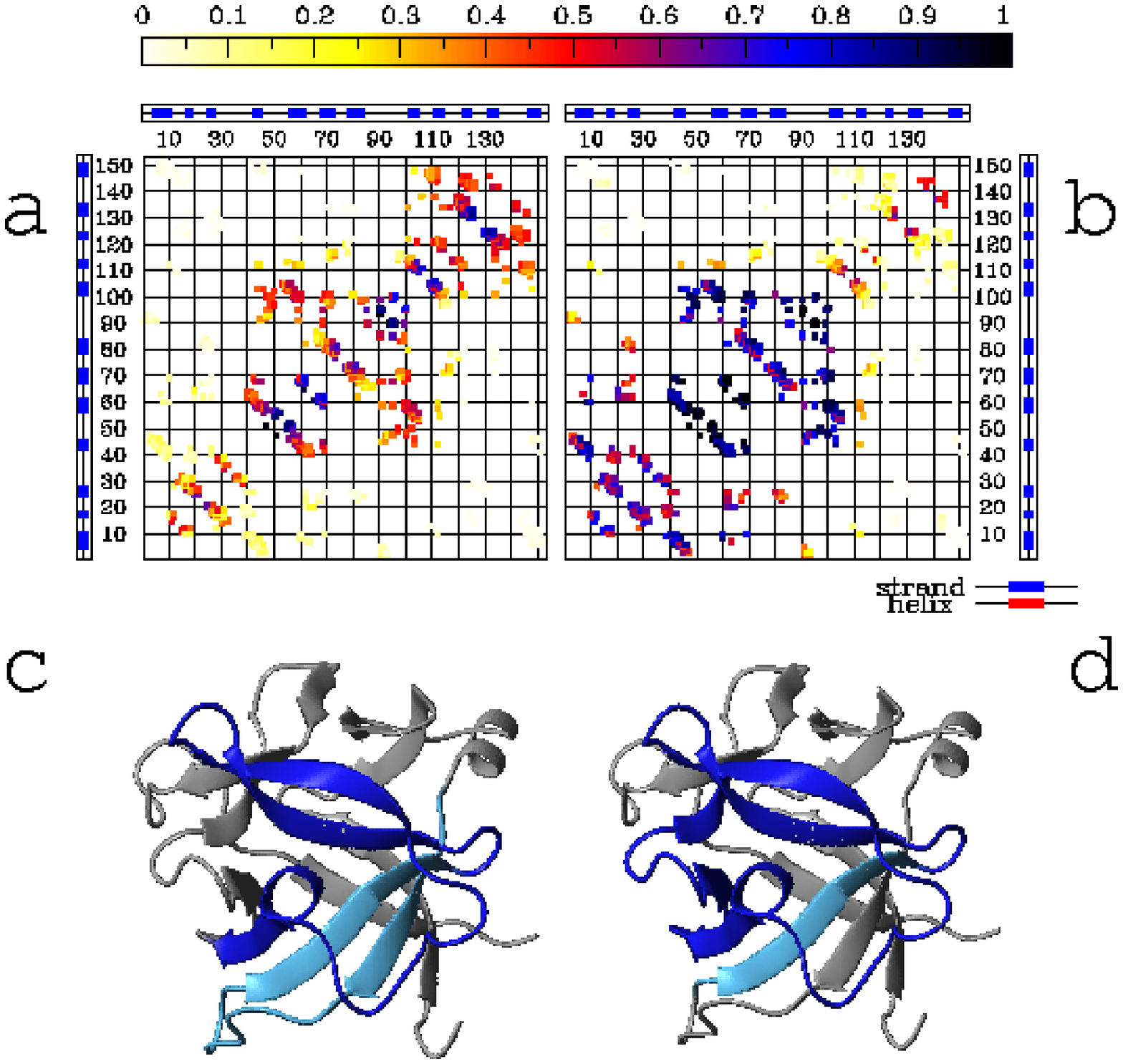,width=0.9\textwidth,clip=}}
\caption{
}
\label{fig4:fig}
\end{figure}

\vspace{2truecm}
{\bf Fig. 2}
The probability  $Q_{ij}(Q)$ of the native DHFR 
contacts to be formed, as resulting from the simulations at different 
stages of the folding
process: {\bf (a)} at an early stage ($Q= 0.1 \pm 0.05$), {\bf (b)}
 at the main intermediate -- located in the interval $Q=0.4 \pm 0.05$ (see 
panels (a) and (c) of Fig.~\protect\ref{fig1:fig}) and {\bf (c)}
at a late stage of the folding process ($Q = 0.7 \pm 0.05$). 
In an topologically and energetically perfectly smooth  funnel-like 
energy landscape, at any value
$Q$  during the folding,  any contact $(i,j)$ should have a probability
$Q_{ij}(Q)$ to be formed equal to $Q$ \protect\cite{Onuchic97}.
By computing $Q_{ij}(Q)$ for each
contact over  different windows of the reaction coordinate $Q$, we can
quantify  the deviations from this smooth funnel behavior and locate
the early  and late contacts along the folding process. It is worth
noticing that any deviation from the ``perfectly smooth"
behavior is mainly due to  topological constraints, since energetic 
frustration has been mostly removed from the system. 
Different colors in the contact maps indicate different probability
values from 0 to 1, as quantified by the color scale at the top. 
The preference to form more local structure than non-local in the
almost unfolded state (b) is due to the smaller conformational entropy 
loss by forming local contacts than by pinching off longer loops 
\protect\cite{plotkinnew}.
The most interesting result is that {\it
domain 1}, identified by the interactions among strands 2-5 and
helices  2-3, is substantially formed at the intermediate
$I_{HF}$ (probabilities for individual contacts grater than 0.7),
while the formation of {\it domain 2} (i.e.
interactions among strand 1, strands 6-8, helix 1 and helix 4) is 
highly unfolded (contact probabilities between 0
and 0.4). Helix 1 and helix 4 are largely formed, but their
interactions with the remainder of the proteins are loose (probabilities
less than 0.4). Overall, this description of the structure of the
main intermediate $I_{HF}$ -- {\it domain 1}  almost
formed and {\it domain 2} largely unformed -- is in
agreement with the structure of $I_{HF}$ experimentally observed.
Moreover, the latest events in the folding process (panel (c))
appear to be the formation of interactions between strands 7-8  and
the remainder of the protein. This again has been 
experimentally determined.
Panel {\bf (d)} illustrates the regions of the native structure 
that simulations and experiments agree to indicate as 
formed at the intermediate $I_{HF}$.

\vspace{2truecm}
{\bf Fig. 3}
The RMS distances between  the regions of the DHFR structure 
identified as {\bf (a)} {\it domain 1} and 
{\bf (c)} {\it domain 2 } and their 
corresponding native configurations are plotted
versus the reaction coordinate $Q$.
{\it Domain 1} collapses to a structure
close to its native conformation (RMS distance less than  5 \AA) 
in the early stages of folding, leading to the formation of the main
intermediate $I_{HF}$ (located at $Q$  around 0.4) whereas {\it domain 2} remains
largely unfolded (RMS distance larger  than 15 \AA). 
In the interval of $Q$ from 0.6 to 0.8 there are several possible structures. 
Consistently with the multi-channel folding model proposed from experimental evidences 
\protect\cite{Jennings93a}, from panels (a) and (c) one can propound several 
possible ways to proceed from $I_{HF}$ to the folded state.
In panel {\bf (b)} the fraction of native
contacts formed, $Q$, is plotted versus the simulation time for a region 
of our simulations where the transition from folded to unfolded state is observed
(at a simulation temperature slightly higher than the folding temperature, $T=1.01 T_f$). 
A set of structures close to the native state ($Q$ around 0.7)
is transiently populated.  Different colors
represent different energies of a configuration 
(quantified by the top energy scale), as well as for panels (a) and (c).
	%
	%
{\bf(d)} The kinetics mechanism for the folding of DHFR, proposed on the
basis of experimental results 
\protect\cite{Jennings93a,Jones95,Jones94,Kuwajima91,Jones95b,Heidary99}.
Experimentally a first step of folding is detected as a very rapid
collapse of the unfolded form to the burst--phase intermediate ($I_{BP}$)
which has a significant content of secondary structure. 
The  folding state is reached through four different channel, involving 
the formations of the main intermediate $I_{HF}$. 
$I_{HF}$ is represented as a set of structures $I_1 - I_4$ 
structurally similar to each other but proceeding towards the native state with 
a different rate.
These intermediate structures evolve to the native forms $N_1 - N_4$ 
via slow-folding reactions.

\vspace{2truecm}
{\bf Fig. 4}
	%
Probability of the contact formation for the
native contacts, as obtained during a typical folding simulation 
(data shown are obtained at $T=0.99 T_f$) 
of the IL-1$\beta$ at different stages of the folding: {\bf (a)}
in a range of $Q$ between 0.3 and 0.4
that corresponds to the early
stage of folding leading to the formation of  the intermediate ensemble;
and {\bf (b)} at the intermediate ($Q$ between 0.45 and 0.55).
At the intermediate the interactions involving strands from 4
to 8 are almost completely formed; interactions among strands 1-3 are
likely  formed but interactions between them and the rest of the
protein are loose. Contacts involving strands 9-12 appear weakened and the
interactions between N (residues 1-40) and C terminus (residues 110-153) are
completely unformed.  
Experimental results confirm that strands 6-8 are well folded in the
intermediate state and that strands 4-5 are partially formed.
Panels {\bf (c)} and {\bf (d)} show the regions of the IL-1$\beta$ 
native structure formed at the intermediate, as resulting from 
simulations (c) and experiments (d). The small difference between simulations
and experiments (contact formation in the region between 
residues 110-125) may be due to energetic considerations that are not taken 
into account in the model, as discussed in the text.
In agreement with experimental results, the formation of contacts between 
N and C terminus is not accomplished until the late stage of folding 
-- these contacts are still unformed for $Q =(0.6-0.7)$ (data not shown).

\newpage

\onecolumn

\end{document}